\documentclass[pra,preprint,superscriptaddress,floatfix,showpacs, 11pt]{revtex4-1}
\usepackage{amsmath,amsfonts,amssymb}
\usepackage{graphicx}
\usepackage{xcolor}

\begin{document}
\title{Assessing small accelerations using a bosonic Josephson junction}
\author{Rhombik Roy}
\email{rroy@campus.haifa.ac.il}
\affiliation{Department of Physics, University of Haifa, Haifa 3498838, Israel}
\affiliation{Haifa Research Center for Theoretical Physics and Astrophysics, University of Haifa,
Haifa 3498838, Israel}
\author{Ofir E. Alon}
\affiliation{Department of Physics, University of Haifa, Haifa 3498838, Israel}
\affiliation{Haifa Research Center for Theoretical Physics and Astrophysics, University of Haifa,
Haifa 3498838, Israel}

%\date{\today}

\begin{abstract}
 Bosonic Josephson junctions provide a versatile platform for exploring quantum tunneling and coherence phenomena in ultracold atomic systems. While extensive research has examined the Josephson-junction dynamics in various double-well configurations, most studies have been limited to inertial reference frames.
In the present work, we posed the question how placing a Josephson junction in a non-inertial reference frame would impact the quantum tunnelling. Our findings demonstrate that accelerating a Josephson junction alters the tunneling dynamics. Conversely, tunneling behavior can be used to assess the acceleration of the system. By analyzing the changes in physical properties, we can assess the acceleration of the double-well.
We begin with the most simple non-inertial frame: moving with constant acceleration. The tunneling time decreases exponentially as acceleration increases, making it effective for measuring larger accelerations. However, for smaller accelerations, accurate assessment requires accounting for many-body depletion, which decreases linearly as acceleration rises.
Next, we explore a more complex scenario where the acceleration is time dependent. In this case, the acceleration is mapped onto the tunneling time period and depletion, which again serve as predictors of acceleration.
We go further by conducting a detailed analysis of the change in tunnelling dynamics when the system deviates from constant or zero acceleration. The quantitative analysis show that the depletion changes exponentially near constant acceleration, while around zero acceleration, the change follows a polynomial pattern.
All in all, we quantify how the tunneling process, as well as the mean-field and many-body properties, evolve in a non-inertial system of increasing complexity.
\end{abstract}

\maketitle

\section{Introduction}\label{intro} 
Precise acceleration measurement is essential in diverse research domains, including gravity-field studies, navigation systems, seismology, and robotics. The field of accelerometry has witnessed the emergence of several innovative methodologies for the detection and measurement of acceleration which includes high-Q optical microsphere resonators~\cite{ext2} and superconducting quantum interference devices (SQUIDs) based accelerometers~\cite{ext3}.
Also, atom interferometry has proven to be a valuable instrument for the assessment of acceleration~\cite{acc1,acc2} and rotation~\cite{rot1,rot2}. Recent advancements in atomic accelerometers using optical lattice setups have demonstrated the ability to sense both the magnitude and direction of applied forces~\cite{Metrology10}. 
While various methods are available for precise measurement, quantum metrology has emerged as a powerful approach, offering high-resolution and highly sensitive measurements of physical parameters beyond classical limits~\cite{Metrology1,Metrology2,Metrology3,Metrology4,Metrology5,Metrology6,Metrology7,Metrology8,Metrology9}. 
Studies have shown that the quantum Fisher information of acceleration reaches peak values, indicating optimal estimation precision in these systems~\cite{PhysRevD.101.056014}.
Quantum-optics-based tools, such as Mach-Zehnder interferometers, offer precise gravitational acceleration estimations~\cite{PhysRevA.99.023803}. 
Also, coherent chemical reactions in ultracold Bose-Einstein condensates (BECs) are highly sensitive to external forces, particularly acceleration. Using advanced spectroscopy, these reactions can be precisely detected, making BECs promising for high-sensitivity sensor applications~\cite{ext1}.
Thus, numerous studies have been conducted to quantify acceleration very preciously through the lens of quantum theory, establishing this area as fertile ground for future exploration.\\

Bosonic Josephson junctions (BJJs), commonly implemented through ultracold bosons tunneling within a double-well potential, offer a promising platform for investigating macroscopic quantum phenomena~\cite{intro1}. 
These setups facilitate the exploration of essential quantum phenomena, such as quantum phase transitions~\cite{intro2,intro3}, quantum entanglement~\cite{intro4}, atom interferometers~\cite{intro5}, and macroscopic quantum coherence~\cite{intro6}.
Due to the unprecedented level of control achievable in experiments, direct observations of tunneling have become experimentally feasible for ultracold bosons~\cite{boson3,boson4,boson5,boson6,boson7,boson8} and fermions~\cite{fermion1,fermion2,fermion3,fermion4}. 
Concurrently, the BJJ dynamics have also been extensively studied theoretically. These investigations have delved into a broad spectrum of phenomena, including Josephson oscillations~\cite{tunneling1,tunneling2,tunneling3,tunneling4}, collapse and revival dynamics~\cite{self_trap_r}, self-trapping effects~\cite{self-trap1,rhombik_epjd,self-trap2,self-trap3}, and fragmentation~\cite{mctdhb_exact3,sudip_asymmetric_dw}. 
Tunneling and quantum self-trapping are intriguing phenomena that strongly depend on the interaction strength between particles. These effects have been studied using both short- and long-range interactions~\cite{rhombik_epjd,sudip_longrange_dw,anal_2020_dw}.
Although extensive research has been conducted in this field, most studies have focused on inertial reference frames.\\

The popularity of the BJJ has sparked our interest in exploring its tunneling dynamics, particularly when placed in a non-inertial reference frame. Additionally, we want to explore whether changes in the tunnelling phenomena can shed light on how the junction is moving.
In this study, we investigate how acceleration affects the BJJ dynamics, focusing on two key indicators: the mean-field survival probability and many-body depletion. Our findings show that acceleration significantly influences both of these properties, with depletion dynamics being particularly sensitive to changes in acceleration. These two quantities could serve as an effective tool for measuring acceleration.
We begin with the simplest case of a non-inertial frame with constant acceleration. In this scenario, changes in the tunneling time period and depletion are highly dependent on the magnitude of acceleration and serve as effective indicators of acceleration. 
We observe that tunneling time decreases exponentially, which serves as a good gauge of larger acceleration. On the other hand, depletion is a good predictor for smaller acceleration, showing a linear decay as acceleration increases. 
We expand our analysis to include scenarios where acceleration is time dependent. Our findings also show that the above key measures can effectively quantify these types of accelerations.
We go one step further and analyze small deviations from accelerated motion in two common cases: deviations from constant acceleration and deviations from zero acceleration.  We demonstrate that even in these more intricate scenarios, we can predict small deviations with precision through the depletion measurement. Specifically, the deviations from constant acceleration show exponential behavior, whereas the deviations from zero acceleration follow a polynomial pattern. 
This research not only broadens our understanding of BJJ dynamics in non-inertial frames but also proposes novel approaches for assessing acceleration.

The structure of this paper is as follows: Section \ref{model} introduces our simulation setup and acceleration protocol. In Section \ref{keymeasures}, we discuss the key quantities measured in our study. Section \ref{results} presents the numerical findings and outlines the method for assessing both constant and time-dependent accelerations. Finally, Section \ref{conclusion} summarizes our results and discussions. Additionally, we include three appendices. Appendix \ref{MCTDHB} discusses the numerical method used in the calculations, appendix \ref{convergence} demonstrates the accuracy of our numerical results, and appendix \ref{tilt} establishes a mapping between a constantly accelerated double-well and a tilted double-well at rest.

\section{Setup and protocol} \label{model}
In this study, we investigate an intriguing aspect of quantum mechanics: tunneling dynamics in a double-well (DW) potential. But, we are not looking at this phenomenon in a typical inertial frame of reference. Instead, we are exploring it in a non-inertial frame, which adds an extra layer of complexity and interest to our investigation.
Our setup involves a symmetric DW potential that moves with trajectory $x_{mov}(t) = \frac{1}{2}a_0t^{\alpha}$, where $a_0$ and $\alpha$ are the acceleration parameters that we can adjust to explore different avenues.
In such a scenario, the velocity of the DW is simply $\dot{x}_{mov}(t)= \frac{\alpha}{2} a_0 t^{(\alpha -1)}$
and the acceleration of this motion is given by the second time-derivative: $\ddot{x}_{mov}(t) = \frac{\alpha (\alpha -1)}{2} a_0 t^{(\alpha -2)}$. 
When $\alpha =2$, the potential moves with constant acceleration. The magnitude of acceleration is given by $\ddot{x}_{mov} = a_0$. $\alpha=1$ corresponds to the  DW moving with constant velocity. There is no acceleration in this case. For all other values of $\alpha$, the acceleration becomes time dependent. 
This parametrization allows us to explore a range of scenarios, from constant velocity to constant acceleration to more complex time-dependent accelerations, all by adjusting the value of $\alpha$.

\begin{figure}
    \centering
    \includegraphics[width=0.65\textwidth, angle =-0 ]{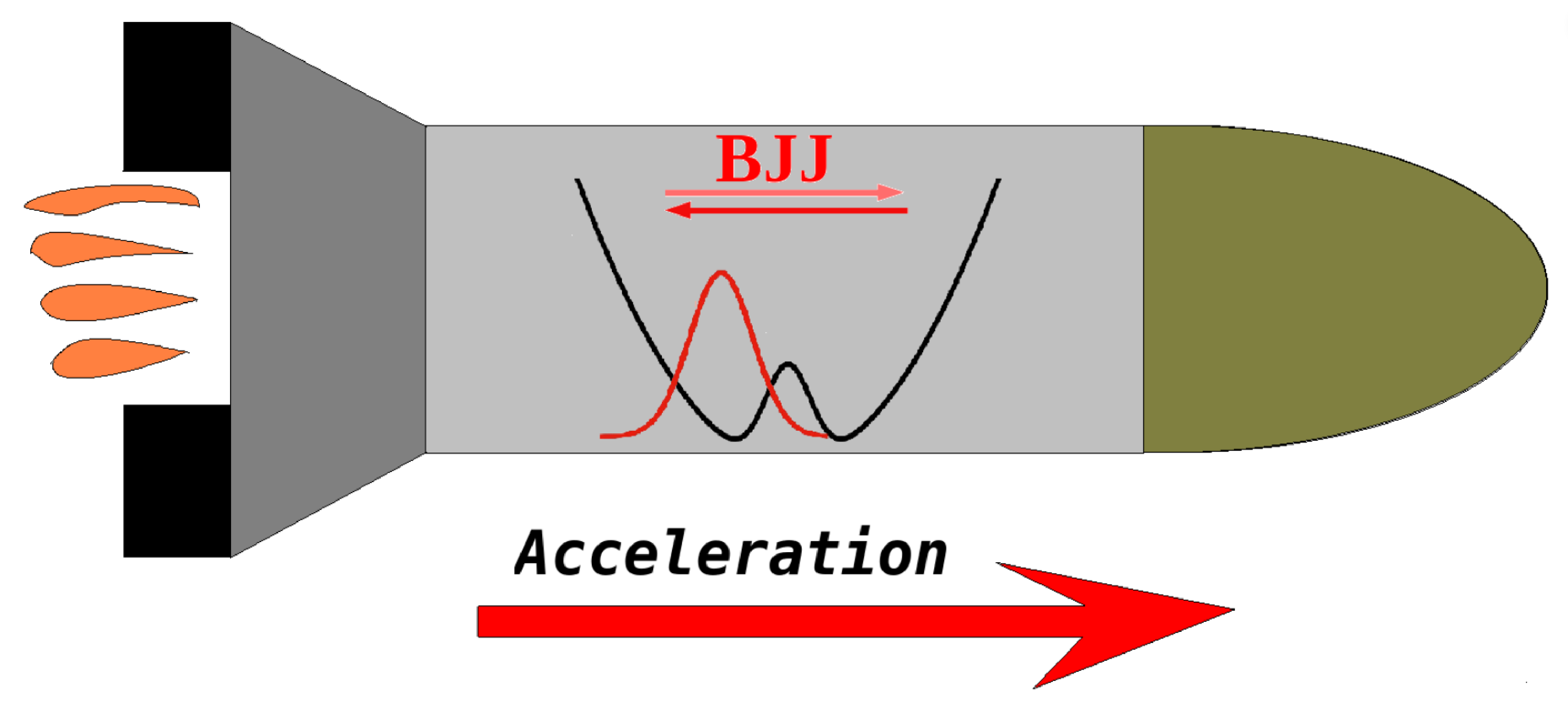}
    \caption{An artistic representation for a bosonic Josephson junction in a non-inertial reference frame.}
    \label{fig0}
\end{figure}

The time evolution of $N$ interacting  bosons is governed by the time-dependent many-body Schr\"odinger equation: $\hat{H} \Psi =i\frac{\partial \Psi}{\partial t}$. The time-dependent Hamiltonian has the form
\begin{equation}
    \hat{H} (x_1,x_2, \dots x_N;t) = \sum_{j=1}^N \hat{h}(x_j;t) + \sum_{k>j=1}^N \hat{W}(x_j-x_k), 
\end{equation}
where $\hat{h}(x;t) = \hat{T}(x) + \hat{V}_{DW}(x;t)$ is the one-body part of the Hamiltonian. $\hat{T}(x)$ is the kinetic energy term %$-\frac{1}{2} \frac{\partial^2}{\partial x^2}$, 
and $\hat{V}_{DW}(x;t)$ is the symmetric double-well trapping potential moving according to $x_{mov}(t)$. $\hat{W}(x_j-x_k)$ is the two-body interaction term in the Hamiltonian.
The bosons interact through a weak repulsive interaction which for convenience we model by a finite-range Gaussian potential of the form $\hat{W}(x_i-x_j) = \frac{\lambda_0}{\sqrt{2 \pi \sigma^2}} e^{\frac{(x_i - x_j)^2}{2\sigma^2}}$, where $\lambda_0$ represents the interaction strength and $\sigma =0.25$~\cite{sudip_asymmetric_dw,sudip_longrange_dw}. The mean-field interaction parameter is chosen as $\Lambda = \lambda_0 (N-1) = 0.1$, with $N$ being the number of particles.
Our computational setup spans a system size from $x_{min} = -8$ to $x_{max} = +242$, discretized into $2048$ grid points. As the DW is in motion, this large range is needed to capture the full dynamics. We have confirmed that our calculations converge with respect to the grid size and density. 
We solve the Schr\"odinger equation by transforming the Hamiltonian $\hat{H}$ into a dimensionless form. This is done by dividing the dimensionful Hamiltonian by $\frac{\hbar^2}{mL^2}$, where $L$ is a convenient length scale and $m$ is the mass of the boson~\cite{MCTDHB2,rhombik_pra}. In this representation, the unit of time is $\frac{mL^2}{\hbar}$.

A well-defined protocol allows for the conversion of dimensionless quantities into their corresponding real, measurable values. This can be illustrated using $^{87}$Rb atoms as an example. The typical size of an experimental double-well setup is in the order of a micron. Thus, choosing $L = 1 \mu m$ for the unit of conversion is convenient. The mass of a $^{87}$Rb atom is $m = 1.4431\times 10^{-25}$ kg.
With these values, we can determine that the unit of time in dimensionful units is $\frac{mL^2}{\hbar} = 1.37$ msec. In comparing dimensionful and dimensionless  units, $1$ sec corresponds to $729.92$ time units, and $1$ m equals $10^6$ length units. For acceleration, the conversion between real and dimensionless units is $1 m/s^2 = \frac{10^6 \text{ (length units)}}{729.92^2 \text{ (time units)}^2} = 1.8769 \text{ dimensionless acceleration units}$. To put this into perspective, a constant acceleration of, for instance, $a_0 = 0.0015$ in the dimensionless units would correspond to $\sim 8\times10^{-4} m/s^2$ in SI units, which is rather very small. This example demonstrates that though we work with convenient dimensionless units in our calculations, one can easily convert them to real units for experimental implementation.

We begin by initializing all bosons on the left side of the DW potential by using the initial trapping potential defined as $V_{initial} = \frac{1}{2}(x+2)^2$. 
At $t > 0$, the bosons are released into the DW potential, which simultaneously begins to accelerate according to the parameters $\alpha$ and $a_0$ as previously defined. 
The form of the DW potential is 
\begin{equation}\label{potential}
    V_{DW}(x)= 
\begin{cases}
    \frac{1}{2}(x+2)^2,& \text{if } x < -\frac{1}{2}\\
    \frac{3}{2}(1-x^2),& \text{if } -\frac{1}{2}\leq x \leq \frac{1}{2}\\
    \frac{1}{2}(x -2)^2,& \text{if } x > \frac{1}{2},
\end{cases}   
\end{equation}
where at $t > 0$, the DW starts to move as $V_{DW}\left(x-x_{mov}(t)\right)$.
Fig.~\ref{fig0} presents an artistic sketch of our numerical simulation setup, where the DW is placed in an accelerating moving object and the tunneling dynamics is investigated through the numerical experiment.

%%%%%%%%%%%%%%%%%%%%%%%%%%%%%%%%%%%%%%%%%%%%%%%%
Our protocol is designed for situations where the initial state starts from rest. However, if the system has an initial velocity $v_0$, this velocity should be accounted for to prevent any abrupt changes in the motion at start.
This could be done by multiplying the initial wave function by a phase factor $e^{+iv_0x}$. This adjustment effectively adds the initial velocity $v_0$ in the positive x-direction to the state that is originally prepared at rest in the initial potential $V_{initial}$. In this scenario, the motion of the accelerated DW is given by $x_{mov}(t) = v_0t + \frac{1}{2}a_0t^\alpha$. 
Of course, when studying the effect of acceleration, the measured quantities remain identical regardless of whether the initial state is prepared at rest or with a constant velocity. However, if there is an initial velocity, the above phase should be applied to adjust for it.
%%%%%%%%%%%%%%%%%%%%%%%%%%%%%%%%%%%%%%%%%%%%%%%%

We use the multiconfigurational time-dependent Hartree method for bosons to solve the $N$-boson Schr\"odinger equation~\cite{MCTDHB1,MCTDHB2}. This method expands the many-body wave function using time-dependent permanents, created by distributing $N$ bosons across $M$ single-particle orbitals~\cite{mctdhb_exact2}. The time-dependent variational principle is employed~\cite{variational5}, leading to two sets of coupled equations that are solved simultaneously to determine the time evolution of the wave function. For a more detailed explanation and applications, see the recent review~\cite{mctdhb_review} and appendix~\ref{MCTDHB}.

\section{Measured quantities}\label{keymeasures}
The main purpose of this work is to assess acceleration with the help of the tunnelling dynamics of bosons in a DW. For this purpose, we accelerate the DW and measure certain physical properties that are affected when the DW is in a non-inertial frame. We study the system both from mean-field and many-body perspectives. 

In order to adequately capture the time evolution, we study the survival probability in the left well. Initially, the DW is placed in such a way that the tip of the barrier lies at $x=0$. The survival probability in the left well is defined as
\begin{equation}\label{eq.survival}
    P_L (t)= \int_{x= -\infty}^{x_{mov}(t)} \frac{\rho(x,t)}{N} dx ,
\end{equation}
where $\rho(x,t)$ is the one-body density. To calculate the survival probability, we use the one-body density obtained from mean-field calculations. This quantity is applicable for large BECs, i.e., BECs with thousands and more atoms.

Generally, quantum correlations build up during tunnelling~\cite{sudip_longrange_dw} and fragmentation manifests itself in correlation functions~\cite{ofir_ref21}. Also, it is now possible to experimentally measure the correlations~\cite{fragmentation_exp1,fragmentation_exp2,fragmentation_exp3}. Consequently, we are particularly interested in how the natural orbital occupations evolve dynamically in an accelerated DW potential. 
The natural occupations are defined as the eigenvalues of the reduced one-body density matrix.
A system is condensed if the first natural orbital is fully occupied $\frac{n_1}{N} \simeq 1$, and fragmented if multiple orbitals are occupied~\cite{frag1,frag2,frag3,frag4,frag5}. During tunneling, correlations develop, causing particles to leave the initially condensed state. The depletion from condensation is defined as
\begin{equation}
    Depletion (\%) = (1-n_1) \times 100.
\end{equation}
In mean-field calculations, only one orbital is considered. Thus, to measure depletion, one needs a many-body treatment. Initially, we prepare a highly condensed state in the left well. 
Depletion is highly sensitive to acceleration, making it an effective tool for precise acceleration measurement, see below.

\begin{figure}
    \centering
    \includegraphics[width=0.55\textwidth, angle =-90 ]{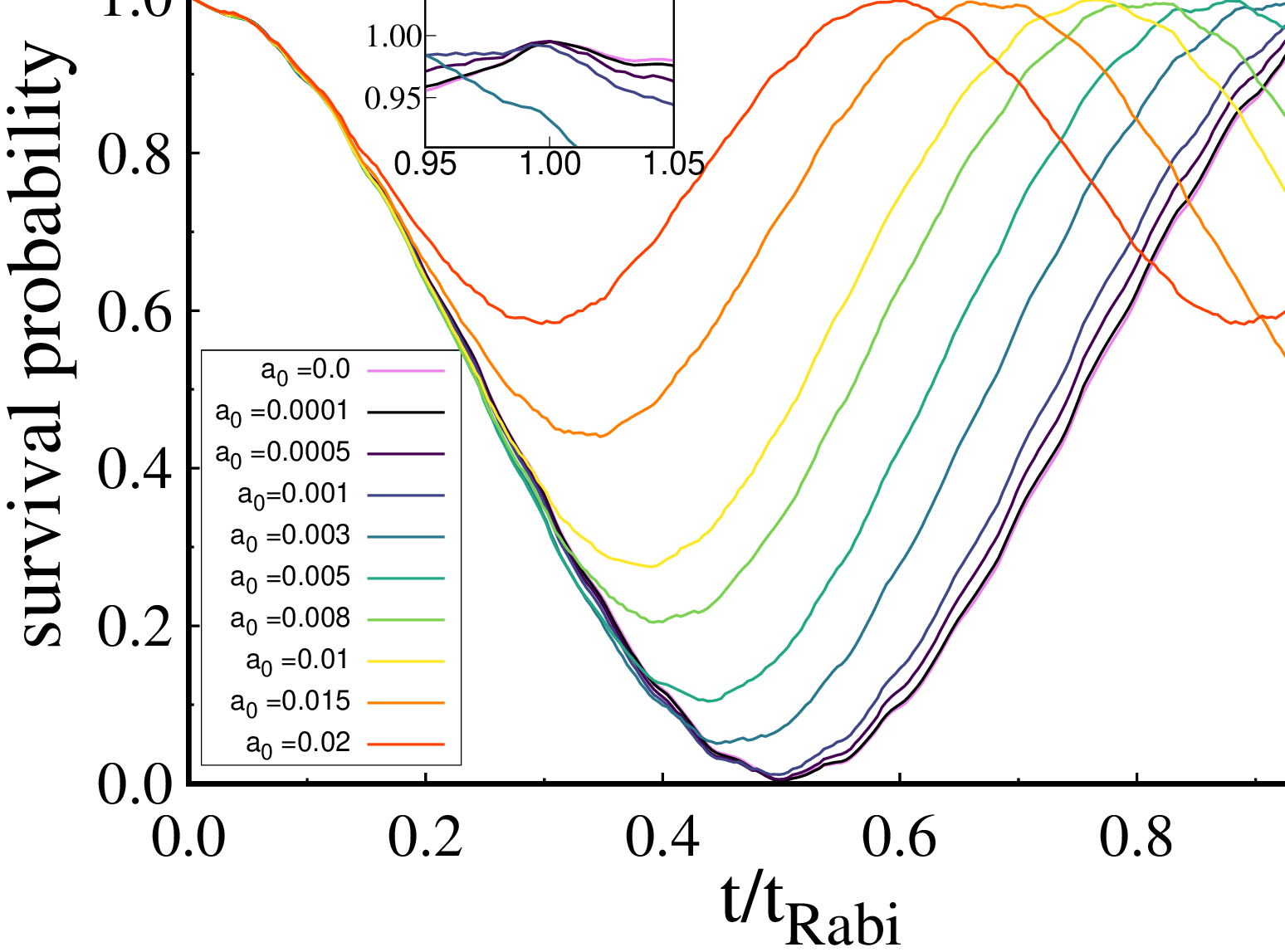}
    \caption{\textbf{Survival probability for constant accelerations.} The survival probability in the left well over one complete Rabi cycle is presented for various accelerations. The interaction parameter is $\Lambda=0.1$. As the acceleration increases, the tunneling time period decreases. For small accelerations, the oscillation period is close to $t_{Rabi}$ (as demonstrated in the inset), leading to difficulties in precisely determining the time period. All quantities are dimensionless.}
    \label{fig1}
\end{figure}

\section{Results and discussion}\label{results}
This section presents a detailed analysis aimed at understanding the tunneling dynamics in an accelerating double-well potential and provides an estimation of the acceleration. Our numerical findings are divided into two separate sections. First, the case of constant acceleration ($\alpha = 2$) is analyzed, with $a_0$ representing the magnitude of acceleration applied to the system. Following this, cases involving time-dependent acceleration, characterized by $\alpha \ne 2$, are examined.
We evaluate acceleration using the time period of the BJJ dynamics and the depletion from the condensate, both of which are significantly influenced by acceleration. Our findings show that changes in the tunneling time period can be used to assess acceleration. As said, here we employ a mean-field calculation. However, this method exhibits limitations in the small acceleration regime. To address this constraint, we propose the use of many-body depletion as an alternative measure, which proves an effective quantity in the small acceleration regime.
For this analysis, we consider a system of $N=10$ bosons.

\subsection{Constant accelerations}
We start with all bosons in the left well of a stationary DW trap. At $t>0$, the bosons are released into the double-well potential as the trap begins to accelerate. For constant acceleration, the motion of the DW is given by $x_{mov}(t) = \frac{1}{2} a_0 t^2$.
Before presenting our main results, it is crucial to note that a symmetric DW moving with constant acceleration can be mapped onto a tilted DW at rest. In Appendix \ref{tilt}, we demonstrate this mapping procedure. This equivalence provides a significant numerical advantage, as it eliminates the need to simulate the actual movement of the DW, which would require a larger system size with a much larger number of grid points.
This mapping allows us to study any magnitude of constant acceleration using the corresponding tilted DW configuration in the lab-frame.

Fig.\ref{fig1} displays the survival probability in the left well for one Rabi cycle for different amplitude of accelerations ($a_0$). The survival probability for the left well is calculated using Eq. \ref{eq.survival} based on the mean-field one-body density. The case where $a_0 = 0.0$ represents a DW at rest. It is worth noting that we have verified that the tunneling dynamics remains identical for both a DW moving at constant velocity and one at rest. 
Throughout our analysis, we use $t_{Rabi}$ to denote the tunneling time period for the DW at rest. We calculate the Rabi oscillation time as $t_{Rabi} = \frac{2 \pi}{\Delta E} = 132.4$, where $\Delta E$ is the energy difference between the ground state and the first excited state of a single particle for the DW at rest.

Fig.~\ref{fig1} demonstrates that as the acceleration increases, the time period of tunnelling dynamics decreases. This is evident since a constantly accelerating double well can be viewed as a tilted double well (see Appendix \ref{tilt}). This tilt naturally results in an inverse relationship between acceleration and tunneling period~\cite{sudip_asymmetric_dw}.
\begin{figure}
    \centering
    \includegraphics[width=0.5\textwidth, angle =-90 ]{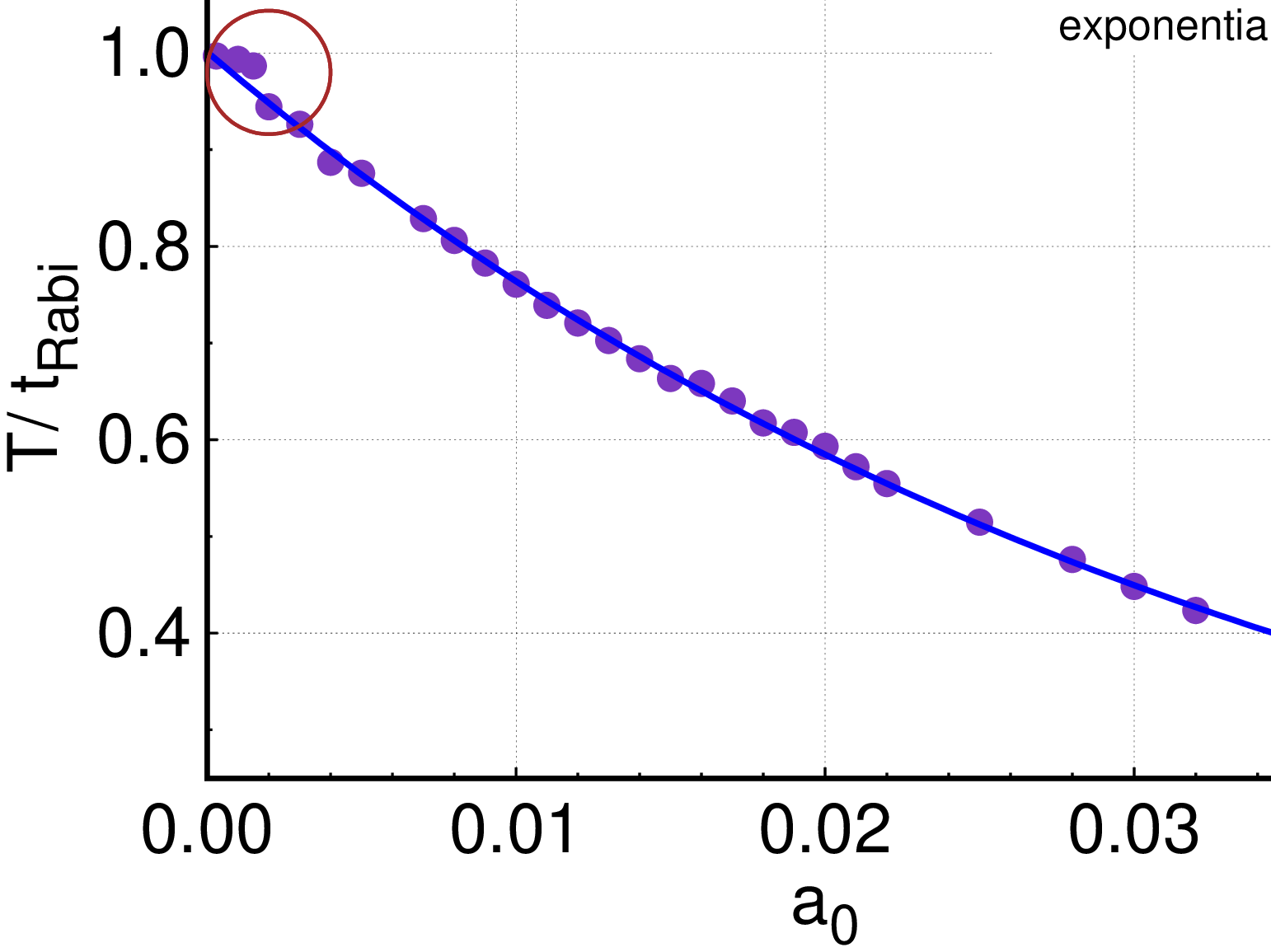}
    \caption{\textbf{Time period as a function of acceleration.}
The relation between the time period and acceleration in an accelerated DW, extracted from Fig.~\ref{fig1}, is presented. The interaction parameter is $\Lambda=0.1$. The time period exhibits an exponential decay with increasing acceleration. However, in the regime of small acceleration, deviations from this exponential behavior are observed, highlighting the limitations of this method in accurately measuring smaller accelerations.  All quantities shown are dimensionless.}
    \label{fig2}
\end{figure}
The tunneling period is measured and plotted as a function of $a_0$ in Fig.~\ref{fig2}. This graph more clearly illustrates the rapid decrease in the tunnelling time period ($T$) with the increase in $a_0$. Our analysis shows that the decay follows an exponential pattern given by $T/t_{Rabi}=0.967 e^{-28.1 a_0} + 0.033$. 
This relation offers a straightforward method for evaluating acceleration through the tunneling period, expressed as $a_0 = -\frac{1}{28.1} \text{ln} \frac{T/t_{Rabi}-0.033}{0.967}$. 
%Given the well-fitted curve, one can determine $a_0$ by observing the time period. 
Importantly, as survival probability is calculated from mean-field density, the behaviour does not depend on the number of bosons used in the numerical analysis. Instead, the dynamics is determined by the mean-field interaction parameter $\Lambda$, which is the product of the interaction strength and the number of particles $N$. So, in the mean-field study, $N$ can be chosen as large as relevant keeping $\Lambda$ fixed.

While effective for detecting larger accelerations, the time period analysis method becomes less reliable for smaller ones. The inset of Fig.~\ref{fig1} shows that for smaller $a_0$ values, the tunneling period is close to $t_{Rabi}$, making precise measurement challenging. Consequently, using the time period to assess the acceleration proves inadequate when the acceleration is very small. 
It is important to note that, throughout the text, small acceleration refers to cases where the acceleration has minimal impact on the tunneling time period, i.e., $T \sim t_{Rabi}$. Conversely, when there is a noticeable difference between $T$ and $t_{Rabi}$, it is considered large acceleration.

\begin{figure}
    \centering
    \includegraphics[width=0.5\textwidth, angle =-90 ]{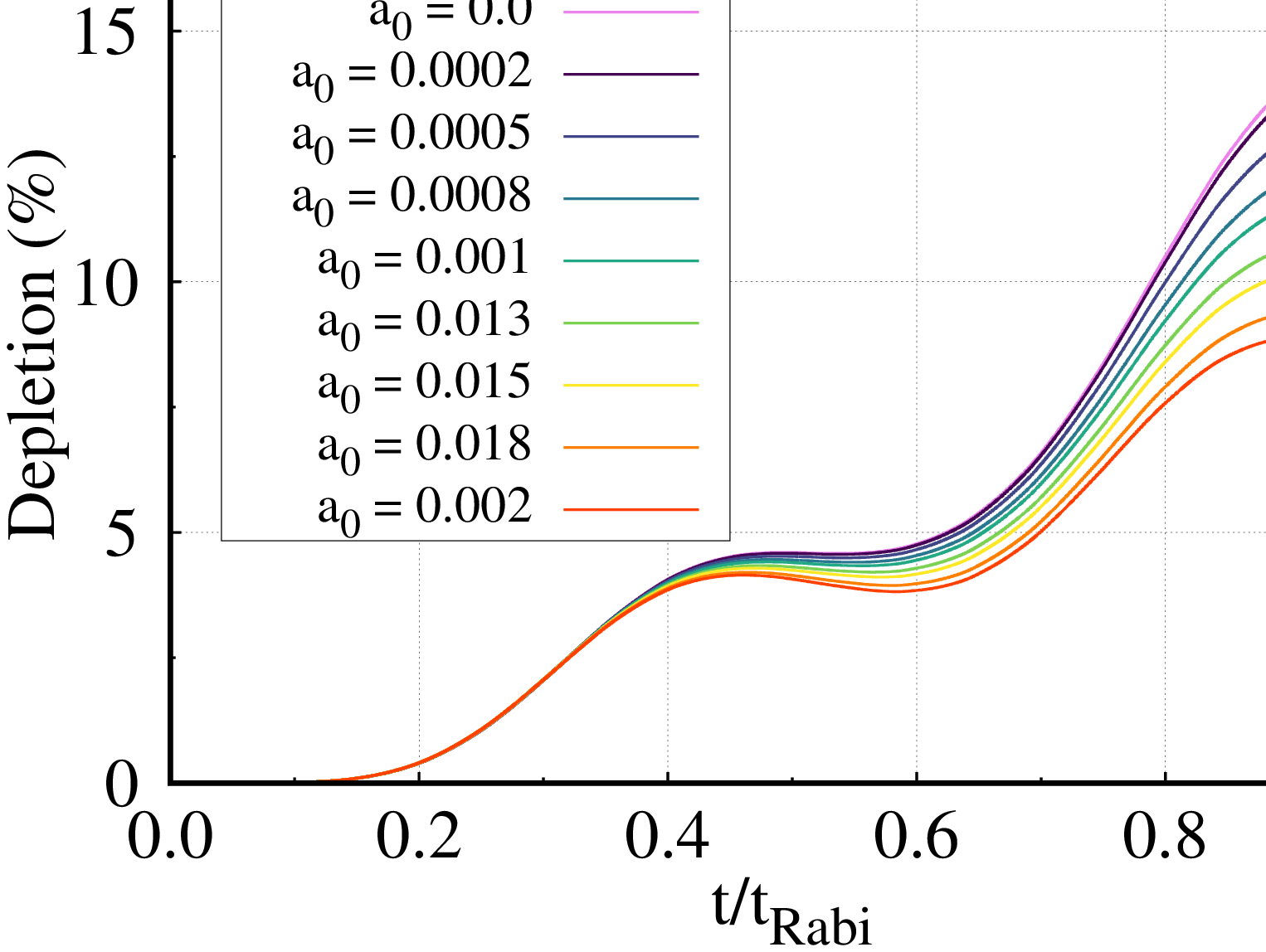}
    \caption{\textbf{Condensate depletion.} The time evolution of many-body depletion is plotted upto one complete Rabi cycle for different accelerations. The higher is the acceleration, the lower is the depletion after one complete cycle. The calculation is performed using $N=10$ bosons and the interaction parameter is $\Lambda=0.1$. The quantities shown are dimensionless.}
    \label{fig3}
\end{figure}

To assess smaller accelerations, we introduce a second quantity---the depletion---which is a many-body phenomenon. 
Before analyzing depletion dynamics, note that the initial depletion at $t = 0$ is about $10^{-3}\%$ because of the weak inter-boson interaction, i.e., we start the dynamics from a highly condensed state.
Fig.~\ref{fig3} illustrates the dynamics of the condensate depletion during tunneling. We present depletion up to one complete Rabi cycle for various accelerations, focusing primarily on the small acceleration regime, since larger accelerations can be detected by measuring the tunneling time period. The figure reveals an inverse relationship between the magnitude of acceleration and the depletion; as the acceleration increases, the number of depleted particles decreases.
This phenomenon can be understood by considering that increasing the acceleration is equivalent to introducing a higher tilt in a fixed DW. Consequently, the initial condensate in the left well tends to remain in the left well, resulting in lower depletion. 

\begin{figure}
    \centering
    \includegraphics[width=0.5\textwidth, angle =-90 ]{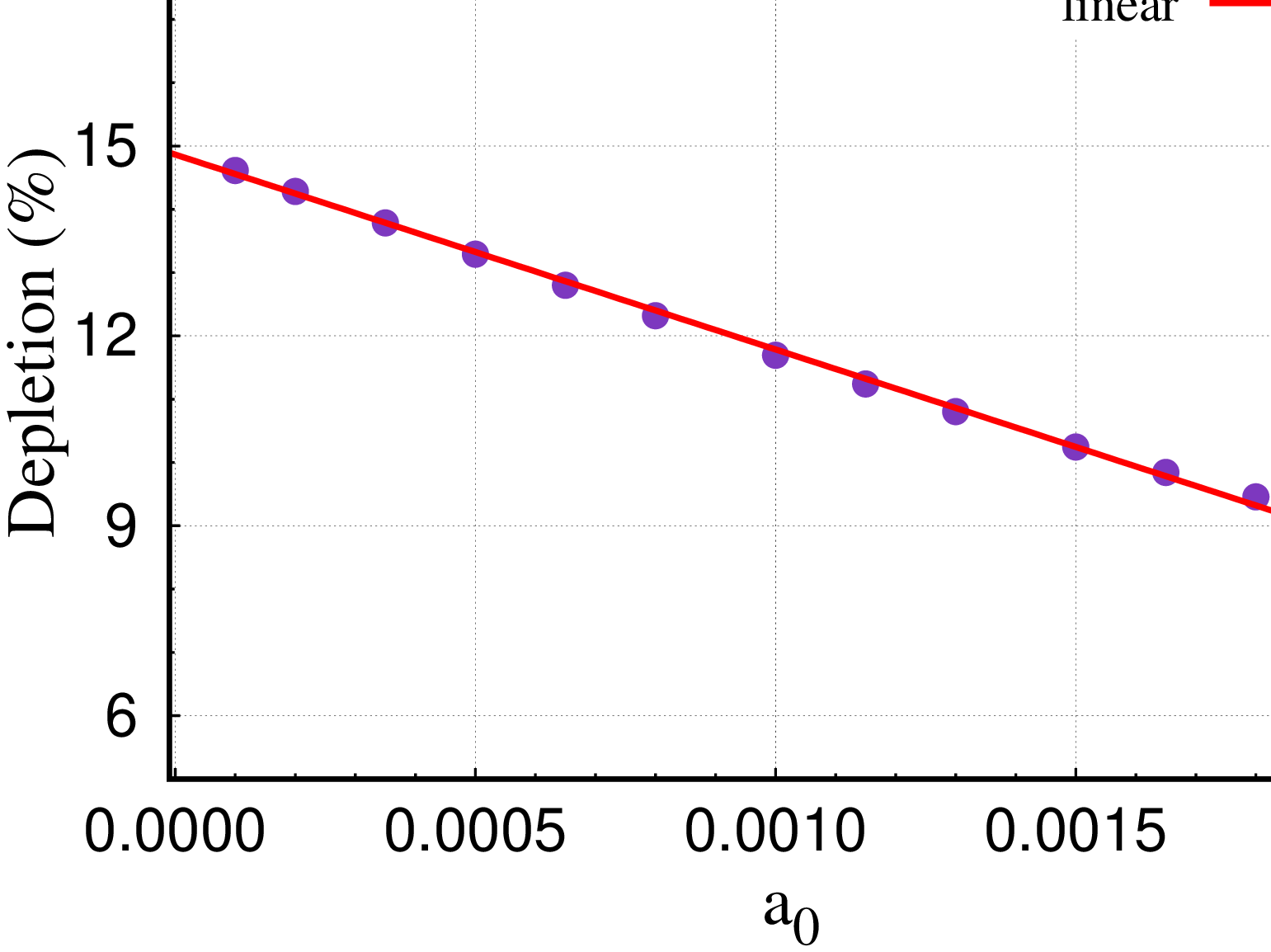}
    \caption{\textbf{ Inferring constant acceleration from depletion.} The graph shows the depletion at $t_{Rabi}$ for different accelerations. The number of bosons is $N=10$, and the interaction parameter is $\Lambda=0.1$. As the acceleration increases, the depletion at $t = t_{Rabi}$ decreases linearly. This linear relation is used to evatuate acceleration from the observed depletion. See text for more details. The quantities shown are dimensionless.}
    \label{fig4}
\end{figure}

A more systematic study of the depletion dynamics, shown in Fig.~\ref{fig4}, allows us to better discuss how to evaluate small accelerations with the help of depletion. In the small acceleration regime, the time period is close to $t_{Rabi}$. We measure the depletion exactly at $t_{Rabi}$ and plot the depletion  as a function of $a_0$ in Fig.~\ref{fig4}. This figure demonstrates that depletion decreases linearly with increasing acceleration, following the fit $Depletion \%=(-30.81 a_0 + 0.1486)\times 100$.  
Using this information, small accelerations can be accurately estimated by measuring the depletion with the formula $a_0 = \frac{14.86 - Depletion\%}{3081}$.
However, increasing acceleration significantly affects the tunneling time period, leading to reduced accuracy in the depletion measurement protocol. Therefore, the many-body depletion method is suitable only for assessing small accelerations, where changes in the time period are comparable to $t_{Rabi}$.

So far, we have two mechanisms for assessing acceleration. The first involves using the mean-field survival probability, which measures the oscillation time period to determine acceleration. However, this method is less effective in the smaller regime. The second method utilizes condensate depletion, which allows for the assessment of smaller accelerations but is less accurate for larger ones. Combining these two key measures allows BJJ dynamics to evaluate a wide range of accelerations, from very small to large. 
Below, we discuss the boundary where we can utilize the time period or the depletion method to assess the acceleration. 
\begin{figure}
    \centering
    \includegraphics[width=0.9\textwidth, angle =-90 ]{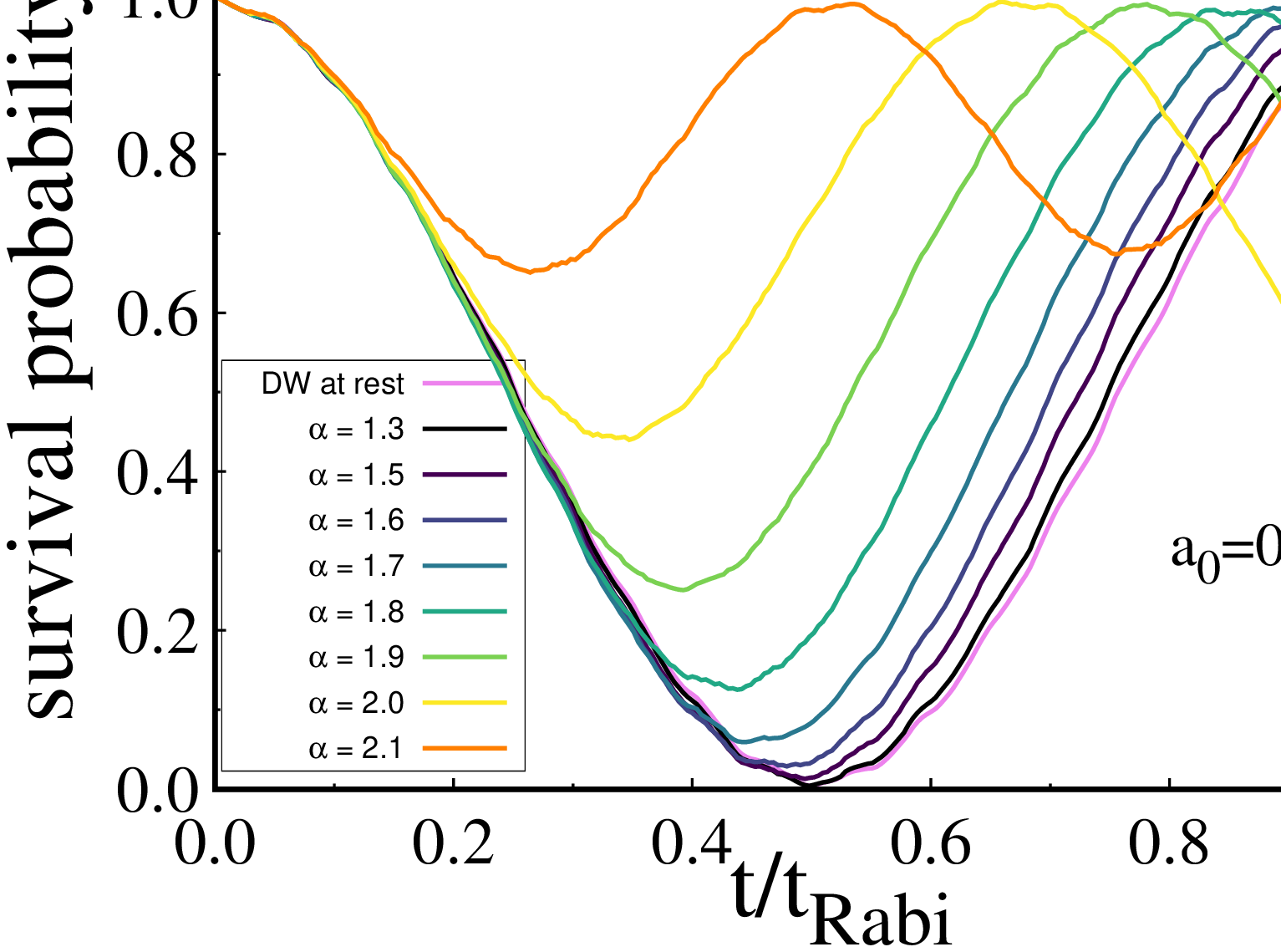}
    \caption{\textbf{Survival probability for time-dependent acceleration.} Mean-field calculations of the survival probability over a complete Rabi cycle are shown for different values of $\alpha$. The interaction parameter is $\Lambda=0.1$. (a) The survival probability is presented for $a_0 = 0.003$. The inset shows that for smaller $\alpha$, the tunneling time period closely matches the $t_{Rabi}$. (b) The survival probability for $a_0 = 0.015$ is depicted. In this case, the tunneling time periods are well-separated due to the five times increase in $a_0$. All quantities are dimensionless.}
    \label{fig5}
\end{figure}

\subsection{Time-dependent accelerations}%%%%%%%%%%
In this section, we examine the effects of time-dependent acceleration on the tunneling time period and depletion.
To this direction, assume that the DW is moving following $x_{mov}(t) = \frac{1}{2} a_0 t^\alpha$, where $\alpha \ne 2$. 
In this case, $a_0$ is, of course, not the acceleration. Rather, the true acceleration is time-dependent and given by $\ddot{x}_{mov}(t) = \frac{\alpha (\alpha -1)}{2} a_0 t^{(\alpha -2)}$. It is important to note that the acceleration at any given time is jointly determined by both parameters, $\alpha$ and $a_0$. 

First, we would like to discuss how the time period of the tunnelling dynamics get affected in the time-dependent acceleration case. Fig.~\ref{fig5} shows the survival probability in the left well up to one complete Rabi cycle. As acceleration is time dependent, we choose two specific $a_0$ values to demonstrate the tunnelling behaviour for different $\alpha$. 
In Fig.~\ref{fig5}(a), the tunneling dynamics for various $\alpha$ over one Rabi cycle is presented for $a_0 = 0.003$. As $\alpha$ increases, the acceleration rises for a fixed $a_0$ at any instant of time, affecting the time period. We observe that higher $\alpha$ values lead to shorter time periods. However, for lower $\alpha$ values, determining the time period becomes challenging as the time periods are close to $t_{Rabi}$ (see inset). In contrast, with $a_0=0.015$, five times larger than before [Fig.~\ref{fig5}(b)], the tunneling dynamics become clearly distinguishable for different $\alpha$ values, making it easy to determine the time period.
We have also confirmed that the tunneling time period decays exponentially with increasing $a_0$ for all $\alpha$ values. 
The yellow line in Fig.~\ref{fig5} represents tunneling dynamics for constant acceleration ($\alpha = 2$). When the DW moves with $\alpha < 2.0$, the time period increases, while for $\alpha > 2.0$ it decreases. This difference becomes more pronounced at higher values of $a_0$. This behavior allows us to easily detect deviations from constant acceleration marking $\alpha=2.0$ as the reference point. However, a problem would arise when the value of $a_0$ is very small. In this case, all the graphs essentially overlap, making distinction difficult. To address this, we need to rely on the depletion dynamics. 

\begin{figure}
    \centering
    \includegraphics[width=0.5\textwidth, angle =-90 ]{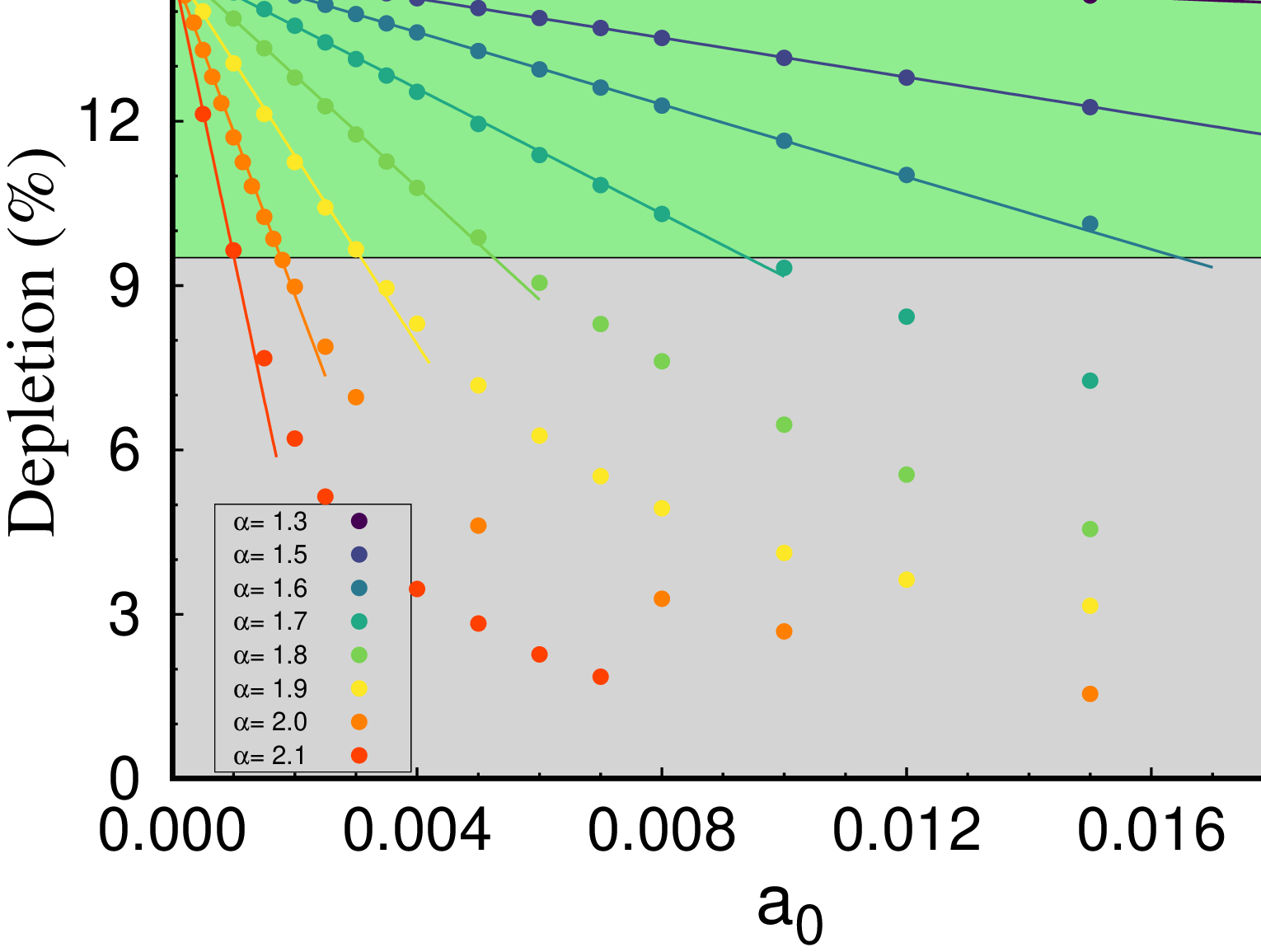}
    \caption{\textbf{Assessing varying accelerations from depletion.} The depletion at $t_{Rabi}$ vs. $a_0$ for different $\alpha$ values is presented. As $a_0$ increases, the depletion at $t = t_{Rabi}$ decreases linearly, with a steeper slope for larger $\alpha$ values. This linear trend can be utilized to estimate acceleration from the observed depletion. For a fixed $\alpha$, the depletion follows a linear decay up to a certain point, after which it deviates from the linear regime. The calculation is performed using $N=10$ bosons and the interaction strength is $\Lambda=0.1$. See text for more details. The quantities shown are dimensionless.}
    \label{fig6}
\end{figure}

For analyzing the depletion dynamics, we utilize the same protocol as before, i.e., the depletion at $t=t_{Rabi}$ is taken for further analysis. Our previous findings in the constant acceleration case indicate that depletion decreases as acceleration increases. The highest depletion of, approximately 15\%, is observed for the DW at rest.
Fig.~\ref{fig6} shows the depletion at $t_{Rabi}$ vs. $a_0$ for various $\alpha$ values. The graph demonstrates that for a given $a_0$ value, an increase in $\alpha$ corresponds to a higher acceleration, resulting in lower depletion. 
In the analysis of the depletion patterns, we observe a consistent trend across different $\alpha$ values. For a fixed $\alpha$, as $a_0$ increases, the depletion decreases linearly up to a certain point and then deviates from linearity, see Fig.~\ref{fig6}. Specifically, we found that this linear relation holds until the depletion declines to approximately 9.5\%.
Our investigation focuses on addressing two key questions: (i) Why does the depletion pattern deviate from linearity in the gray shaded region of Fig.~\ref{fig6}? (ii) Is it possible to determine the $\alpha$ value from the linear regime of the depletion curve? 

To address the first question regarding the deviation from linearity in the gray shaded region, we need to consider our method of data collection. We measure the depletion value at $t_{Rabi}$, but this approach becomes inadequate at higher accelerations, where the actual tunnelling time period deviates significantly from $t_{Rabi}$. 
The deviation from linearity in the depletion curve of Fig.~\ref{fig6} indicates a significant difference between the actual tunneling time period and $t_{Rabi}$. Therefore, we should shift our focus to directly analyzing the time period in order to accurately assess accelerations in this regime.
We observed that when the depletion decreases by more than 35\% compared to the depletion at $t_{Rabi}$ for a static DW, there is a noticeable deviation in the time period from $t_{Rabi}$. This 35\% threshold indicates the point at which our initial method of using depletion at $t_{Rabi}$ becomes less reliable, necessitating direct time period analysis to determine the acceleration parameters.

To address the second question regarding how to determine $\alpha$ from the linear regime of the depletion curve, we analyze the linear fits of the observed depletion data, best described by $\left(Depletion\% = M a_0 + 14.86\right)$. We found the slopes $M$ to be $-44$, $-179$, $-330$, $-570$, $-1023$, $-1721$, $-3081$, and $-5258$ for $\alpha = 1.3, 1.5, 1.6, 1.7, 1.8, 1.9, 2.0, 2.1$, respectively. The slope ($M$) of the linear fit is observed to follow an exponential relationship with $\alpha$ as $M \sim A e^{5.43 \alpha}$, where $A$ is a constant.
This relationship allows us to estimate $\alpha$ for a given $a_0$, provided that the depletion stays within the 35\% bounds shown in the green shaded region of Fig.~\ref{fig6}.

\begin{figure}
    \centering
    \includegraphics[width=0.55\textwidth, angle =-90 ]{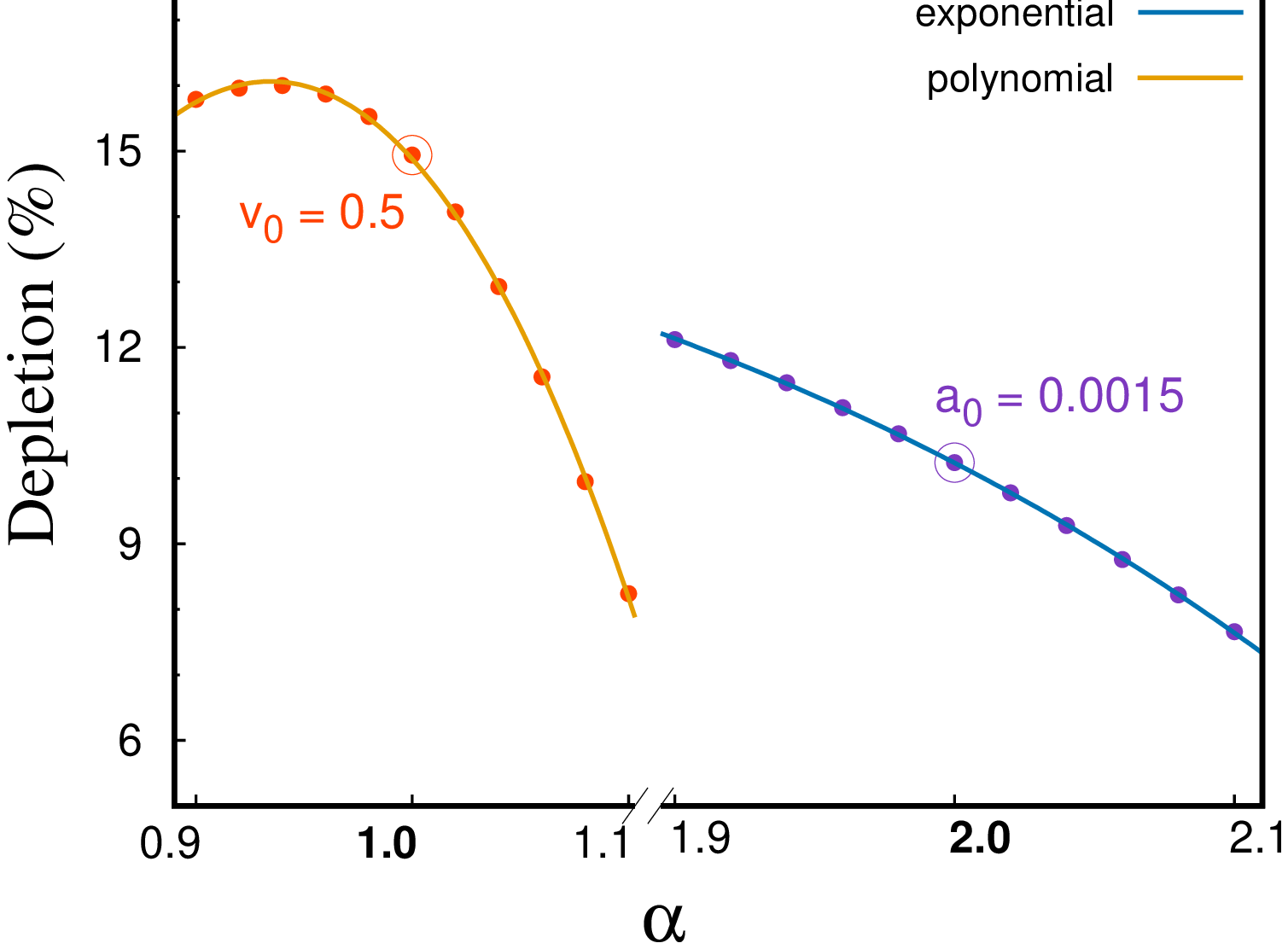}
    \caption{\textbf{Assessing small deviations from constant velocity and constantly accelerated motions.} This figure illustrates how depletion changes when the motion of an vessel deviates from the expected trajectories of constant velocity (orange double circle) and constant acceleration (purple double circle). For $\alpha$ values other than 1 or 2, the acceleration is time-dependent. In the regime around constant acceleration ($1.9 < \alpha < 2.1$), the depletion decreases exponentially with increasing $\alpha$. In the regime around constant velocity ($0.9 < \alpha < 1.1$), the depletion follows a third-order polynomial fit. Calculations were performed using $N = 10$ bosons and the interaction parameter is $\Lambda=0.1$; see the main text for further details. All quantities shown are dimensionless.}
    \label{fig8}
\end{figure}
%%%%%%%%%%%%%%%%%%%%%%%%%%%%%%%%%%%%%%%%%%%%%%%

Up to this point, our investigation has focused on two main aspects. First, the tunneling behavior of the DW under constant acceleration was examined. Next, we have explored time-dependent acceleration scenarios ($\alpha \ne 2$). In both cases, methods to measure the acceleration are discussed.
Next, we aim to explore a more realistic scenario. Consider a vessel programmed to move with either constant velocity or constant acceleration. However, due to a malfunction in the system, the vessel is not following the intended motion. Our goal is to detect the deviation from these expected movements. Specifically, we focus on cases where the $\alpha$ values are close to, but not exactly, $2$ or $1$. Since the focus is on detecting small deviations, only the many-body depletion is discussed in these cases.

We begin by exploring the case where the motion deviates from constant acceleration. We analyze the impact of this deviation on the tunneling dynamics. Fig.~\ref{fig8} (purple dots) shows the depletion percentage for various $\alpha$ values, with $a_0 = 0.0015$ selected for example. To detect small deviations, $\alpha$ is selected in the range $1.9<\alpha<2.1$. When $\alpha \ne 2$, the acceleration becomes time-dependent, leading to changes in the depletion.
In the figure, the depletion for constant acceleration ($\alpha = 2$) is marked with a double circle. The results indicate that the depletion increases (decreases) for $\alpha < 2$ ($\alpha > 2$). The depletion value at $\alpha = 2$ (constant acceleration) serves as a reference for assessing the motion of the DW. In the given range, we find that the depletion decreases exponentially as $\alpha$ increases. Therefore, by measuring the depletion, the $\alpha$ value can be easily determined, and deviations from constant acceleration can be assessed.

Secondly, let us consider a vessel moving at a constant velocity, which, by definition, means there is no acceleration acting on it. But the vessel deviates from its ideal motion of constant velocity because of some external factors.
We choose the motion in this case to be described by $x_{\text{mov}}(t) = v_0 t^\alpha$, where $\alpha$ deviates from the ideal value of $1$.
We are interested in understanding how this deviation impacts the depletion dynamics and hence in determining the value of $\alpha$.
Given that the DW has an initial velocity $v_0$, we account for this by multiplying the initial wave function prepared in the left well by $e^{+iv_0x}$. To measure small changes, we focus on the depletion for $0.9 < \alpha < 1.1$ in Fig.~\ref{fig8} (orange dots). 
The DW is considered to move with a constant velocity of $v_0 = 0.5$, with the corresponding depletion marked by a double circle in Fig.~\ref{fig8} ($\alpha=1$). 
The relation between the depletion and $\alpha$ follows polynomial fit in this regime. 
The case $\alpha > 1$ corresponds to accelerating system where the depletion decreases as $\alpha$ increases. The case $\alpha<1$ corresponds to deceleration, causing the depletion to initially increase and then decrease as $\alpha$ is reduced from $1$. The behavior of the tunneling process under deceleration is not well understood, making it an intriguing area for theoretical developments and further investigations. 
 
Overall, the BJJ can serve as an effective detector for measuring constant and time-dependent accelerations, as well as deviations from uniform motion.
%%%%%%%%%%%%%%%%%%%%%%%%%%%%%%%%%%%%%%%%%%%%%%%

\section{Conclusions}\label{conclusion}
The development of acceleration measurement techniques has constantly changed our understanding of how particles move and interact, from the tiniest subatomic particles to the largest cosmic structures.
Simultaneously, the study of tunneling dynamics in a DW has emerged as a powerful tool for investigating quantum phenomena across various scales. 
In this work, we merge these two active areas of study by examining the tunnelling dynamics in an non-inertial frame. Specifically, we focus on the tunneling behavior of bosons in a DW potential that is subject to acceleration. This approach allows us to explore how acceleration affects quantum tunneling processes and potentially offers new insights into  acceleration detection. 

Our research uncovers an intricate interplay between the motion of the DW in a non-inertial reference frame and quantum tunneling. We demonstrate that accelerating a DW significantly alters the tunneling dynamics. More intriguingly, we find that these changes in the tunneling behavior can be used to detect and measure the motion of the system itself.
Thus, we propose using a BJJ as a tool for assessing acceleration. Our method relies on measuring two key quantities: survival probability and depletion. For larger accelerations, the effects are readily observable through the decay in the time period of the tunnelling dynamics. On the other hand, depletion, which is a many-body effect and sensitive to the motion, is useful for accurately measuring small accelerations.
We begin our analysis with the simplest non-inertial reference frame, i.e., a system which moves with constant acceleration. In this scenario, the tunneling time period decreases exponentially, while the depletion decreases linearly as the magnitude of the acceleration increases. Building on these findings, we can accurately predict the value of constant acceleration using both mean-field and many-body measures.
The analysis is expanded to include more complex non-inertial frames where acceleration is time dependent. We provide a detailed discussion on the effects of non-constant acceleration on the tunneling time and depletion. Our findings demonstrate that acceleration in these more complex non-inertial frames can also be detected using the measured quantities. 
Furthermore, we examine small changes in accelerating motion in two typical cases: deviation from constant acceleration and from zero acceleration. Even in these complex cases, the many-body depletion allows us to quantitatively predict these small changes. Our findings show that deviations from constant acceleration lead to an exponential pattern in the depletion, while small changes from constant velocity follow a polynomial trajectory.
The immediate open question in this direction is to explore how tunneling is affected by acceleration in higher-dimensional DW potentials. This will be addressed in future work.
 
\section*{Acknowledgements}
This work is supported by the Israel Science Foundation (ISF) grant no. 1516/19. RR acknowledges Anal Bhowmik for helpful discussions. Computation time at the High-Performance Computing Center Stuttgart (HLRS) and High Performance Computing system Hive of the Faculty of Natural Sciences at University of Haifa are gratefully acknowledged.

\appendix

\section{The multiconfigurational time-dependent Hartree for bosons method (MCTDHB)} \label{MCTDHB}
In this Appendix, we discuss the many-body numerical method to solve the time-dependent Schr\"odinger equation. In the MCTDHB framework, the many-body wave function is expanded as a linear combination involving time-dependent permanents as 
\begin{eqnarray}
\vert \Psi(t)\rangle = \sum_{n} C_{n}(t)\vert n;t\rangle; \hspace{5ex}
\vert n;t\rangle = \prod_{i=1}^{M}
 \frac{ \left( b_{i}^{\dagger}(t) \right)^{n_{i}} } {\sqrt{n_{i}!}} \vert vac \rangle .
\label{many_body_wf}
\end{eqnarray}
The system comprises of $M$ orthonormal time-dependent single-particle states, denoted by $\phi_j (x,t) =\langle x\vert \hat{b}_j^{\dagger}(t) \vert vac \rangle$, 
where $ b_{j}^{\dagger}(t)$ is the creation operator. By distributing $N$ bosons among $M$ orbitals, the number of permanents is given by $ \left(\begin{array}{c} N+M-1 \\ N \end{array}\right)$. 
$n = (n_1,n_2, \dots ,n_M)$ represents the occupation of the orbitals and $n_1 + n_2 + \dots +n_M = N$ is the total number of particles~\cite{MCTDHB1,MCTDHB2}.
The set $\vert n;t \rangle$ spans the entire Hilbert space as $M \rightarrow \infty$, and the wave function becomes exact. 
For computational purposes, we must limit $M$ to a finite value. The chosen value of $M$ should be large enough to ensure numerical convergence of our results. We discuss numerical convergence in Appendix~\ref{convergence}, where we analyze how the results vary with different values of $M$. This examination of convergence by varying the number of orbitals is a necessary step to validate our calculation.

In the MCTDHB method, both the expansion coefficients $C_{n}(t)$ and the orbitals $\phi_j(x,t), j = 1, \dots, M$ are time-dependent and fully optimized using the variational principle. This time-dependence allows one to achieve a given level of accuracy with a much shorter expansion compared to methods using time-independent basis sets~\cite{mctdhb_review}.
To obtain the many-body wave function $\vert \Psi (t) \rangle$, we evaluate the time-dependent coefficients $C_{n}(t)$ and orbitals $\phi_j (x,t), j=1, \dots,M$. The time evolution follows a coupled set of first-order integro-differential equations for the orbitals and the coefficients, which are solved simultaneously~\cite{MCTDHB1,mctdhb_software1,mctdhb_software2}. 
MCTDHB has been successfully benchmarked against analytically solvable many-body models~\cite{mctdhb_review}, demonstrating its capability to handle highly correlated problems with time-dependent potentials, and has been widely used in different theoretical calculations~\cite{paolo_ol,fischer_Metrology,mctdhb15,mctdhb17,rhombik_pre,mctdhb21,rhombik_jpb,paolo_cavity}. 
We utilize the numerical tools available in the MCTDH-X package~\cite{mctdhb_software1,mctdhb_software2}.

\begin{figure}
    \centering
    \includegraphics[width=0.8\textwidth, angle =-90 ]{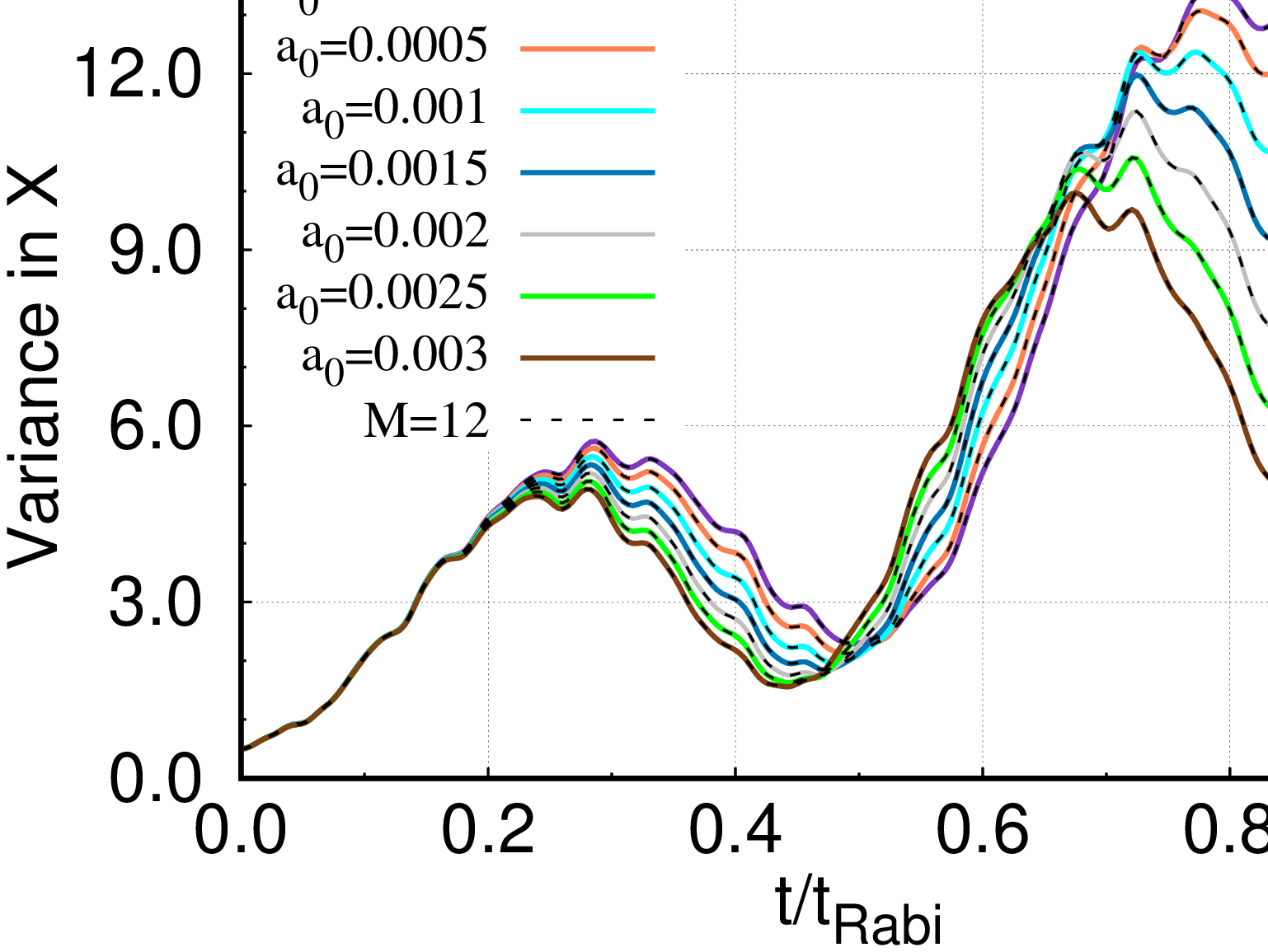}
    \caption{\textbf{Convergence of numerical results for constant accelerations.} (a) The depletion is calculated with $M=8$ and $M=12$ time adaptive orbitals for different accelerations. (b) The convergence of the time-dependent position variance per particle is observed for $M=8$ and $M=12$ time-adaptive orbitals. In both panels, $M=12$ time-adaptive orbital calculations are illustrated by black dotted lines. Irrespective of the magnitude of the acceleration in our calculation, this figure strongly supports the convergence of our results with $M=8$ orbitals. The number of bosons is $N=10$ and the interaction parameter $\Lambda=0.1$. The color codes are explained in each panel. The quantities shown are dimensionless.}
    \label{fig:app_conv}
\end{figure}

\section{Convergence of numerical results} \label{convergence}
This appendix focuses on demonstrating the convergence of the quantities discussed in the results section. The MCTDHB ansatz becomes numerically exact as the number of orbitals approaches infinity, as mentioned in Appendix~\ref{MCTDHB}. However, in most cases, we can accurately describe both ground-state and dynamics using a finite number of orbitals. In dynamical studies, the number of orbitals required to achieve converged results depends on the duration of the dynamics being simulated~\cite{mctdhb_review}.
To determine the number of orbitals necessary for accurate results, a convergence check is essential.
To ensure convergence, we compare the measured quantities calculated with $M$ orbitals to those calculated with $M+4$ orbitals. This comparison helps us to determine the minimal number of orbitals required to accurately describe the dynamics. 
The many-body dynamics presented in Sec.~\ref{results} is numerically calculated using $M=8$ orbitals. 
We assess convergence by comparing the depletion and the sensitive position variance computed with $M=8$ and $M=12$ time-adaptive orbitals. Results are presented for constant acceleration ($\alpha=2.0$) and one time-dependent case ($\alpha=1.9$).

Before analyzing the convergence of the dynamics, it is needed to ensure convergence of the initial condition. To recall, the number of bosons is $N=10$ and the interaction parameter $\Lambda=0.1$. We prepare the initial state in the left well and calculate the energy per particle using different numbers of orbitals. For $M=8$ orbitals, the energy per particle is $E_{(M=8)}=0.51923817$, while for $M=12$ orbitals, it is $E_{(M=12)}=0.51923813$. Additionally, we measure the initial condensate depletion, which is $0.001377\%$ for $M=8$ orbitals and $0.001378\%$ for $M=12$ orbitals. These results demonstrate a very good convergence for the initial condition with $M=8$ orbitals, which is a prerequisite for achieving convergence in the dynamics.

The convergence of the depletion dynamics is illustrated in Fig.~\ref{fig:app_conv}(a). In our study, depletion serves as a valuable many-body property, enabling us to detect smaller accelerations. 
For constant acceleration, we plot the condensate depletion up to one complete Rabi cycle for different accelerations. The solid lines represent calculations using $M=8$ orbitals, while the dotted lines show results for $M=12$ orbitals. As seen in the plot, the results for $M=8$ and $M=12$ orbitals exactly fall on top of each other, confirming the convergence of the depletion dynamics.
\begin{figure}
    \centering
    \includegraphics[width=0.8\textwidth, angle =-90 ]{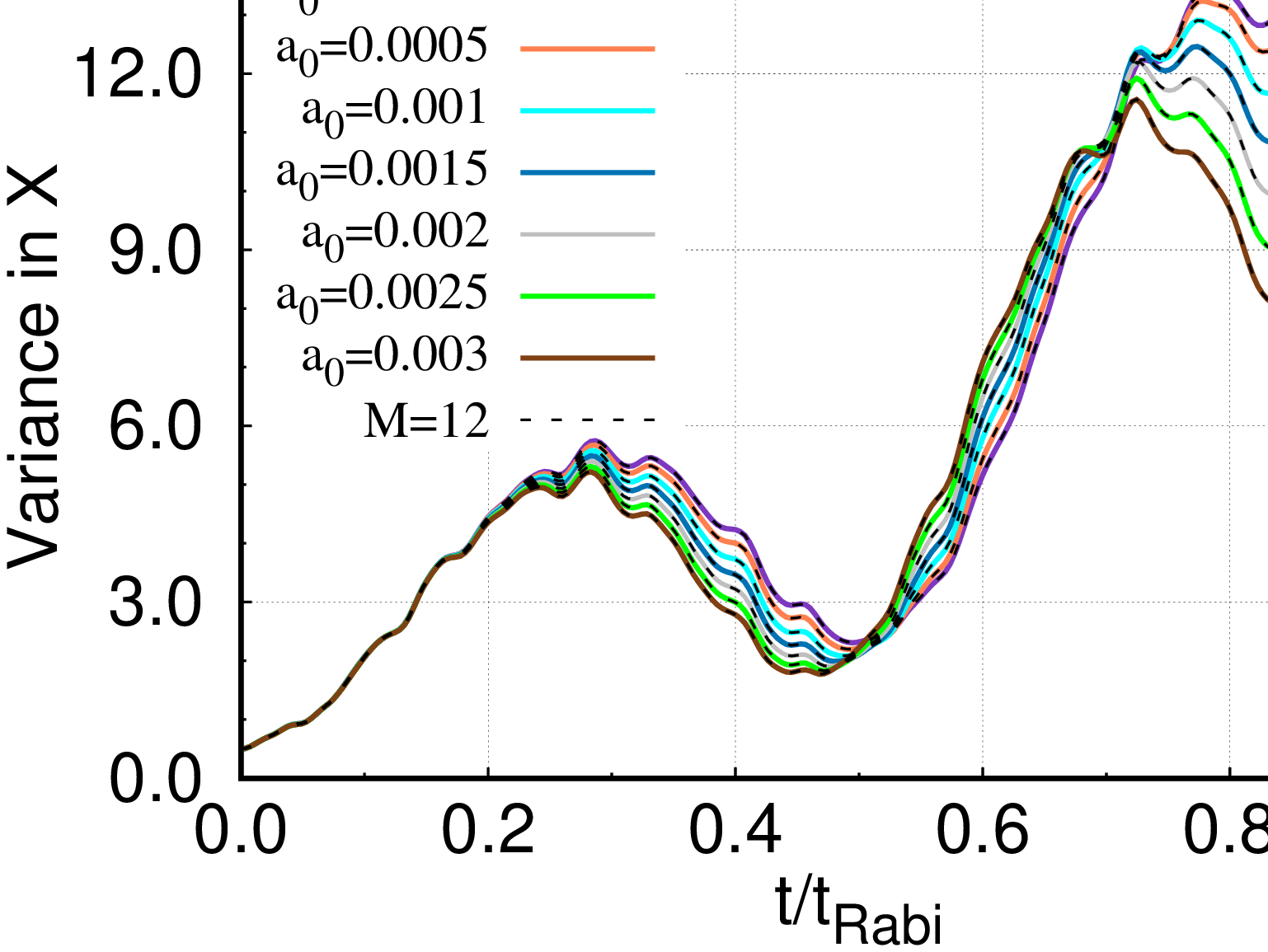}
    \caption{\textbf{Convergence of numerical results for time-dependent accelerations.} (a) The depletion is calculated with $M=8$ and $M=12$ time adaptive orbitals for different accelerations. (b) The convergence of the time-dependent position variance per particle is observed for $M=8$ and $M=12$ time-adaptive orbitals. $\alpha=1.9$ is chosen to illustrate the convergence for the cases where the acceleration is time dependent. In both panels, $M=12$ time-adaptive orbital calculations are illustrated by black dotted lines. Irrespective of the magnitude of the acceleration in our calculation, this figure strongly supports the convergence of our results with $M=8$ orbitals. The number of bosons is $N=10$ and the interaction parameter $\Lambda=0.1$. The color codes are explained in each panel. The quantities shown are dimensionless.}
    \label{fig:app_conv1.9}
\end{figure}

The many-body variance is another useful and sensitive quantity for demonstrating convergence. For this, the many-body position operator in the $x$ direction is given by $\hat{X} = \sum_{j=1}^N \hat{x}_j$. The variance of the many-particle position operator is expressed as:
\begin{equation}\label{eq.variance1}
    \frac{1}{N}\Delta_{\hat{X}}^2 = \frac{1}{N} \left( \langle\hat{X}^2 \rangle - \langle\hat{X} \rangle^2 \right),
\end{equation}
where $\hat{X}^2 = \sum_{j=1}^N \hat{x}^2 (\mathbf{r}_j) + 2\sum_{j<k}^N \hat{x} (\mathbf{r}_j) \hat{x} (\mathbf{x}_k)$, consisting of one-body and two-body operators. Calculating variances is a powerful method for exploring correlations within the system. For detailed discussions, see Refs.~\cite{mctdhb_review,uncertainty_product_alon}.
To show the convergence, Fig.~\ref{fig:app_conv}(b) presents the position variance calculations using the same parameters as before.
Similar to our analysis of the depletion dynamics, we compare position variances calculated with $M=8$ and $M=12$ orbitals. The results overlap precisely, further confirming the convergence of our numerical calculations.
This convergence test using the position variance, a quantity which is particularly sensitive to many-body effects, provides strong evidence that the $M=8$ orbital calculations very accurately capture the dynamics. 

Finally, in Figs.~\ref{fig:app_conv1.9}(a) and~\ref{fig:app_conv1.9}(b), we present the convergence for time-dependent accelerations, using $\alpha = 1.9$ as an example. We display the depletion and position variances for various values of $a_0$ to illustrate this convergence. In both figures, the solid lines represent calculations with $M=8$ orbitals, and the dotted black line represents $M=12$ orbitals. 
The depletion and position variances show that the results overlap well in this case as well. 
These convergence analyses confirm that the many-body results presented in the main text for both constant and time-dependent accelerations, calculated with $M=8$ orbitals, are converged.

\section{Constant acceleration vs. tilted double-well} \label{tilt}
In this appendix, we show the theoretical derivation that a uniformly accelerating symmetric DW can be mapped with a tilted DW at rest. Obviously, the bosons trapped in an accelerated DW experience a pseudo-force due to the DW being in a non-inertial frame. 
We load the bosons in the left well of the DW and it is accelerating in the right direction (see Fig.~\ref{fig0}), then the pseudo-force acts on the bosons in the opposite direction, which pushes the bosons to remain in the left well. So, the left well is energetically favourable. 
This can be imagined as a tilted DW potential where the left side is lower than the right side.
To determine the amount of tilt required, we need to transform the Hamiltonian in the following way.

\begin{figure}
    \centering
    \includegraphics[width=0.63\textwidth, angle =-90 ]{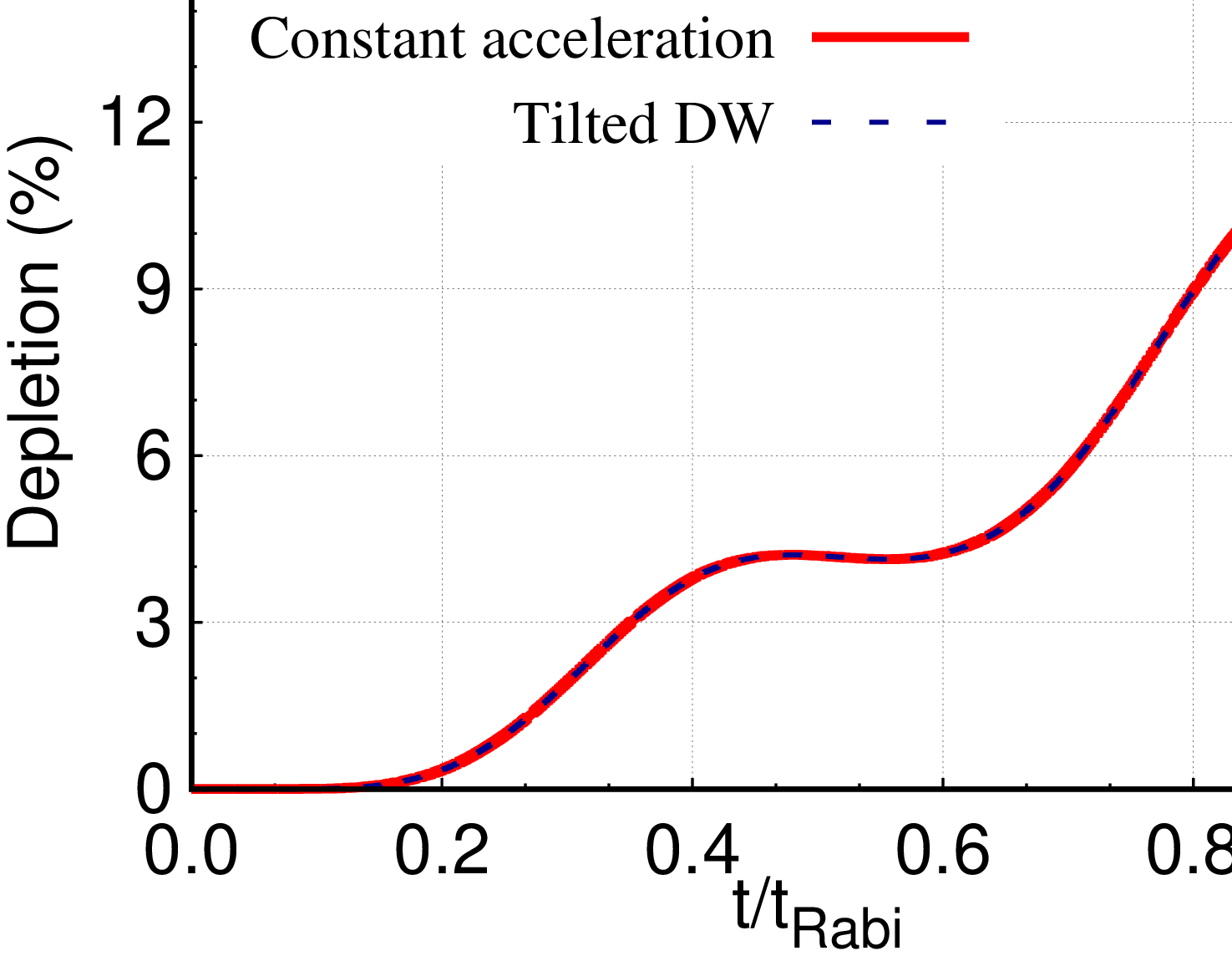}
    \caption{\textbf{Constant acceleration vs. tilted double-well.} Survival probability upto one complete Rabi cycle is presented (a) in mean-field and (b) in many-body calculations. (c) Depletion per particle is presented up to one complete Rabi cycle. (d) Position variance is presented both in mean-field and many-body level. Comparison is made for a static double well with a constant tilt (dotted line) and double-well moving with constant acceleration $a_0 = 0.001$ (solid line). All quantities fall on top of each other. These results align with the analytical calculation that a tilted DW can produce the same dynamics as a symmetric DW under acceleration. The quantities shown are dimensionless.}
    \label{fig:acc_tilt}
\end{figure}

We solve the Schr\"odinger equation $\hat{H}(t)\Psi(\mathbf{x};t) = i\frac{\partial \Psi(\mathbf{x};t)}{\partial t}$, where the full many-body wave function $\Psi(\mathbf{x};t)$ depends on the spatial coordinates $x_1,x_2,\dots ,x_N$ of the $N$ bosons.
The time dependent Hamiltonian has the form $\hat{H}(t) = \sum_{j=1}^N \left[\frac{\hat{p_j}^2}{2m} + V_{DW}(x_j - x_{mov}(t)) + \frac{1}{2} \sum_{k \ne j}^N \hat{W}(x_j - x_k) \right]$, where $x_{mov}(t) = \frac{1}{2}a_0 t^2$ and $\dot{x}_{mov}(t) = a_0 t$. Let us introduce a transformed wave function and write $\Psi(\mathbf{x},t) = e^{-i\sum_l \hat{p}_l x_{mov}(t)} \Phi(\mathbf{x},t)$. 
Note that the two-body interaction term, $\hat{W}(x_j - x_k)$, depends only on the distance between two bosons and remains unchanged under the transformations discussed below. 
Substituting the transformed wave function into the Schr\"odinger equation one has,
\begin{align} 
    \sum_j \left[\frac{\hat{p}_j^2}{2m} + V_{DW}(x_j - x_{mov}(t)) + \frac{1}{2} \sum_{k \ne j}^N \hat{W}(x_j - x_k) \right] e^{-i \sum_l \hat{p}_l x_{mov}(t)} \Phi(\mathbf{x},t) &= \nonumber \\ i\frac{\partial}{\partial t}\left[  e^{-i \sum_l \hat{p}_j x_{mov}(t)} \Phi(\mathbf{x},t) \right] & , \\ 
    \sum_j \left[ \frac{\hat{p}_j^2}{2m} + V_{DW}(x_j) - \hat{p}_j a_0 t + \frac{1}{2} \sum_{k \ne j}^N \hat{W}(x_j - x_k) \right] \Phi(\mathbf{x},t) = i \frac{\partial \Phi(\mathbf{x},t)}{\partial t} & . \label{c2}
\end{align}

Rearranging Eq.~(\ref{c2}), the intermediate result is
\begin{eqnarray} \label{rrrr}
    \sum_j \left[ \frac{1}{2m} \left( \hat{p}_j - m a_0 t\right)^2 - \frac{m^2a_0^2t^2}{2m} + V_{DW}(x_j) + \frac{1}{2} \sum_{k \ne j}^N \hat{W}(x_j - x_k) \right] \Phi(\mathbf{x},t) = i \frac{\partial \Phi(\mathbf{x}, t)}{\partial t}.
\end{eqnarray}
Let us make another transformation to omit the $ma_0 t$ from the first term in Eq.~\ref{rrrr}. The transformation implies $\Phi (\mathbf{x},t) = e^{i\sum_l m x_l a_0t} \xi (\mathbf{x},t)$. Substituting to Eq.~\ref{rrrr} leads to
\begin{align}
     \sum_j \left[ \frac{1}{2m} \left( \hat{p}_j - m a_0 t\right)^2 - \frac{m^2a_0^2t^2}{2m} + V_{DW}(x_j) + \frac{1}{2} \sum_{k \ne j}^N \hat{W}(x_j - x_k) \right] e^{i\sum_l m x_l a_0t} \xi  (\mathbf{x},t) &= \nonumber \\  i \frac{\partial }{\partial t} e^{i\sum_l m x_l a_0t} & \xi (\mathbf{x},t), \\
    \sum_j \left[ \frac{\hat{p}_j^2}{2m}  - \frac{m a_0^2t^2}{2}  +\left\{ V_{DW}(x_j) + m a_0 x_j \right\}  + \frac{1}{2} \sum_{k \ne j}^N \hat{W}(x_j - x_k) \right] \xi (\mathbf{x}, t) = i \frac{\partial \xi(\mathbf{x}, t)}{\partial t} & .     
\end{align}
The term $(-\frac{m a_0^2t^2}{2})$ is the only time-dependent part, and one can eliminate it by doing another transformation. So, after doing the two transformations, we arrive at the modified inertial frame potential as $\left\{V_{DW}(x) + m a_0 x \right\}$, corresponding to a DW moving with constant acceleration $a_0$. So, the symmetric DW potential moving with a constant acceleration $a_0$ can be replaced by an $m a_0 x$ amount of tilt in the DW in the inertial frame.

To demonstrate this equivalence numerically and as another consistency check of our code, we choose the acceleration of, for instance, $a_0 = 0.001$. In the symmetric DW at rest, as described by Eq.~\ref{potential}, we add another term $+0.001x$ to tilt the DW.
We calculate the survival probability, depletion, and position variance both at the many-body and mean-field levels of theory.
Figs.~\ref{fig:acc_tilt}(a) and~\ref{fig:acc_tilt}(b) present the survival probability in the left well for the mean-field and many-body calculations. The solid line corresponds to the survival probability when the DW is accelerating with $a_0=0.001$ and the dotted line is when the DW is tilted by $0.001 x$ amount. In both cases, the dynamics of the survival probability exactly fall on top of each other. Similarly, the depletion [Fig.~\ref{fig:acc_tilt}(c)] and the position variance [Fig.~\ref{fig:acc_tilt}(d)], which is a sensitive quantity, also fall on top of each other for accelerating and tilted DWs. Thus, through numerical and analytical calculations, we show that the dynamics of bosons in a constantly accelerated DW is exactly the same as in a tilted DW.

It is important to note that when the acceleration is not constant ($\alpha \ne 2$), two distinct scenarios arise. For $\alpha > 2$, a transformation similar to the one described above is possible, which maps the time-dependent acceleration to a time-dependent tilt in a double-well potential originally at rest.
In contrast, for $\alpha <2$, an exact mapping is not feasible due to the discontinuity of $\dot{x}_{mov}(t)$ at $t=0$. Consequently, simulations must be conducted directly in an accelerating DW potential.

\end{document}